\documentclass[10pt]{article}
\usepackage{amsmath}
\usepackage{amsbsy}
\usepackage{graphicx}
\usepackage{float}
\usepackage[affil-it]{authblk}

\newcommand{\IM}{\operatorname{Im}}

\title{The photoelectric effect in external fields}

\author[1]{Christian Bracher}
\author[1]{John B.\ Delos}
\author[2]{Vassiliki Kanellopoulos}
\author[2]{Manfred Kleber}
\author[2]{Tobias Kramer}
\affil[1]{Physics Department, The College of William and Mary\\
Williamsburg, VA 23187--8795, USA}
\affil[2]{Physik Department T30, Technische Universit\"at M\"unchen\\
James-Franck-Str., 85747 Garching, Germany}

\date{June 1, 2005}

\begin{document}

\maketitle

\begin{abstract}
Atoms and negative ions interacting with laser photons yield a coherent source of photoelectrons. Applying external fields to photoelectrons gives rise to interesting and valuable interference phenomena. We analyze the spatial distribution of the photocurrent using elementary quantum methods. The photoelectric effect is shown to be an interesting example for the use of coherent particle sources in quantum mechanics.
\end{abstract}

\section{Introduction}

The wave--particle duality lies at the heart of quantum mechanics. Duality applies also to light whose nature had occupied the minds of scientists for centuries. Let us only mention the important milestones: In 1675 Isaac Newton had the idea that light was a stream of tiny particles \cite{Newton1730a} whereas in 1678 his rival Christian Huygens suggested that light behaves like traveling waves. The matter should have been settled in favor of the wave hypothesis with Thomas Young's seminal double--slit experiment (1801) \cite{Young1802a} (but it took another 30 years before British scientists would concede the point). However, in 1900 Max Planck found the famous radiation law \cite{Planck1900a} that describes quantitatively the intensity of light emitted at different frequencies $\nu$ from a hot blackbody. To derive the spectral energy density Planck had to assume that the energy of light ``is composed of a definite number of finite parts'' (which Einstein later called energy quanta) 
\begin{equation} 
E_\nu= N h\nu. 
\end{equation} 
Planck was not sure about the meaning of this equation. He tended to believe that the division of radiation into small portions $h\nu$ was not a property of the radiation field itself but resulted from the interaction between light and matter in thermal equilibrium. He later said ``I can characterize the whole procedure as an act of despair, since by nature I am peaceable, and opposed to doubtful adventures.  However, I had already fought for six years \ldots without arriving at any successful result.  I was aware that this problem was of fundamental inportance \ldots hence a theoretical interpretation \textit{had} to be found at any price, no matter how high it may be'' \cite{Mehra}.

At this point Albert Einstein took action. The title of his famous paper \cite{Einstein1905a} reads \textit{``\"Uber einen die Erzeugung und Verwandlung des Lichtes betreffenden heuristischen Gesichtspunkt.''} [Concerning an Heuristic Point of View Toward the Emission and Transformation of Light]. In his paper Einstein introduced independent energy quanta (the term photon was coined later) which characterize radiation and cannot be divided further. Einstein used a statistical--thermodynamic argument for the existence of such quanta: Examining a formula he derived for the entropy of radiation, he found that high frequency (or low density) radiation ``behaves, in a thermodynamic sense, as if it consists of mutually independent radiation quanta'' of magnitude $h\nu$, and therefore ``it is plausible to investigate whether the laws of the creation and transformation of light are so constituted as if light consisted of such quanta'' \cite{Mehra,Pais1982a}.

In the last three sections of his 1905 paper \cite{Einstein1905a} Einstein's applied the concept of independent energy quanta to explain Stokes' law of fluorescence (light emitted has a lower frequency than the light absorbed), the ionization of gases by ultraviolet light, and the photoelectric effect in solids. He predicted that the maximum kinetic energy of electrons released from the solid would be
\begin{equation}  
\label{eq:wf}
K_{\mathrm{max}} = h\nu - \phi,
\end{equation}
where $\phi$ denotes the work function of the solid. At the time he was only able to state that this formula was ``not in contradiction'' with the available experiments.

Einstein himself said that his theory was ``very revolutionary,'' and indeed it was too much even for his admirers.  Later, when Planck nominated him to the Prussian Academy of Sciences, he felt he had to apologize:  ``There is hardly one among the great problems \ldots to which Einstein has not made an important contribution.  That he may sometimes have missed the target in his speculations, as, for example, in his hypothesis of light quanta, cannot really be held too much against him, for it is not possible to introduce fundamentally new ideas, even in the most exact sciences, without occasionally taking a risk.''

The photoelectric theory was finally put to the test a decade later by Robert Millikan, who showed that the formula was accurate to about 0.5~\%. Still, the hypothesis of light quanta was so incredible to him that he said: ``Despite the apparently complete success of the Einstein equation, the physical theory of which it is designed to be the symbolic expression is found so untenable that Einstein himself, I believe, no longer holds it'' \cite{Millikan}.

Today the quantum theory of radiation is well established.  However we now know that the photoelectric effect is actually not a compelling argument for the existence of photons, because the Einstein relation can be derived from the Schr\"odinger equation using time--dependent perturbation theory \cite{Hall1936a}, assuming a non--quantized, classical electromagnetic wave of frequency $\nu$. Phenomena that truly cannot be explained in terms of classical, non--quantized radiation fields include squeezing via nonlinear optical processes, or the generation of non--classical light \cite{Scully1997a}. Recent beam--splitter experiments with single photons \cite{Hong1986a,Legero2004a} exemplify Einstein's idea of the existence of indivisible photons.

\section{Photoelectric effect as a two--step process}

\begin{figure}[t]
\begin{center}
\includegraphics[width=0.3\textwidth]{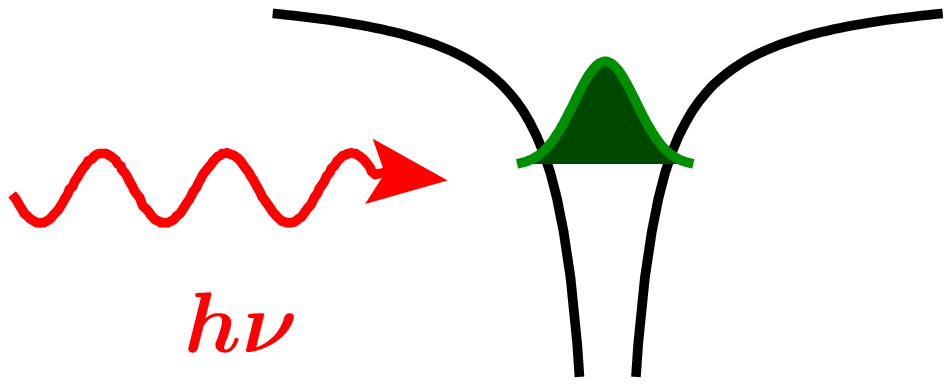}
\includegraphics[width=0.3\textwidth]{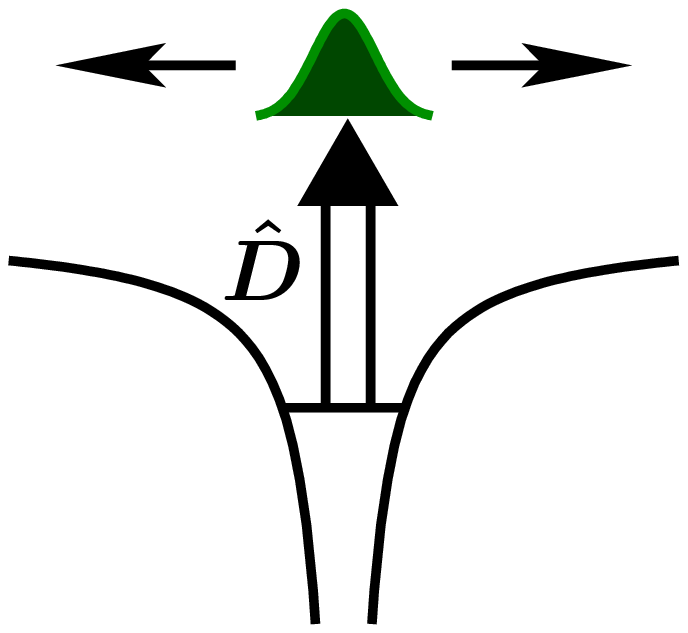}
\end{center}
\caption{Two steps to create a photoelectron: (left panel) the photon transfers its energy $h\protect\nu$ to the initially bound electron, (right panel) the photoelectron escapes from the absorption region and propagates in the applied fields.}
\label{fig:2s}
\end{figure}

New knowledge is presently being gained by examining the photoelectric effect in applied static electric or magnetic fields.  We may speak about this as a two-step process (Fig.~\ref{fig:2s}). In the first step, the incoming photon transfers its energy to the bound electron and a photoelectron is created with an energy $E$ given by Eq.~(\ref{eq:wf}).  In the second step it leaves the atom and propagates in the applied fields. 
The dynamics of the emitted electron is governed by the rules of quantum mechanics. In the following we consider a dilute gas of independent atoms where the interaction of the photoelectron with neighboring atoms can be neglected. Then, under steady--state conditions with many atoms and monochromatic light we can calculate the photocurrent from Fermi's Golden Rule \cite{Heller1978a}: 
\begin{equation}  
\label{eq:FGR}
J(E) \propto \,\,\IM\,\langle\hat{D}\psi|\hat{G}(E)|\hat{D}\psi\rangle,
\end{equation}
with the proportionality factor suppressed. Here $\hat{D}|\psi\rangle$ describes the action of the dipole operator $\hat{D}=\hat{\boldsymbol{\epsilon}}\cdot\mathbf{r}$ on the initially bound state $|\psi\rangle$ of the photoelectron under consideration. $\hat{G}$ is the quantum propagator (or single particle Green function) at fixed energy $E=h\nu-E_0$ (for atoms the work function must be replaced by the positive binding energy $E_0$ of the photoelectron).  This formula says that
the moving photoelectron has to be propagated with the Hamiltonian in the applied fields, and then matched with itself again \cite{Bracher2003a}.
\begin{equation}
\langle\hat{D}\psi|\hat{G}(E)|\hat{D}\psi\rangle = \iint\mathrm{d}\mathbf{r}
\,\mathrm{d}\mathbf{r}^{\prime}\, \langle\hat{D}\psi|\mathbf{r}\rangle
\langle\mathbf{r}|\hat{G}(E)|\mathbf{r}^{\prime}\rangle \langle\mathbf{r}^{\prime}|\hat{D}\psi\rangle.
\end{equation}
The energy--dependent Green function 
\begin{equation}
G(\mathbf{r},\mathbf{r}^{\prime};E)=\langle\mathbf{r}|\hat{G}(E)|\mathbf{r}^{\prime}\rangle
\end{equation}
is the relative probability amplitude for an electron to travel from $\mathbf{r}$ to $\mathbf{r}^{\prime}$. This formulation emphasizes the dynamical aspects of the propagation and opens the possibility of a semiclassical calculation of photocurrents with closed--orbit theories \cite{Kleppner2001a,Du1988a}. In general there are several classical orbits that link the points $\mathbf{r}$ and $\mathbf{r}^{\prime}$. In quantum mechanics, these paths have to be weighted with complex amplitudes and coherently summed up \cite{Berry1972a}.

\section{Near--threshold effects}

\begin{figure}[tbp]
\begin{center}
\includegraphics[width=0.4\textwidth]{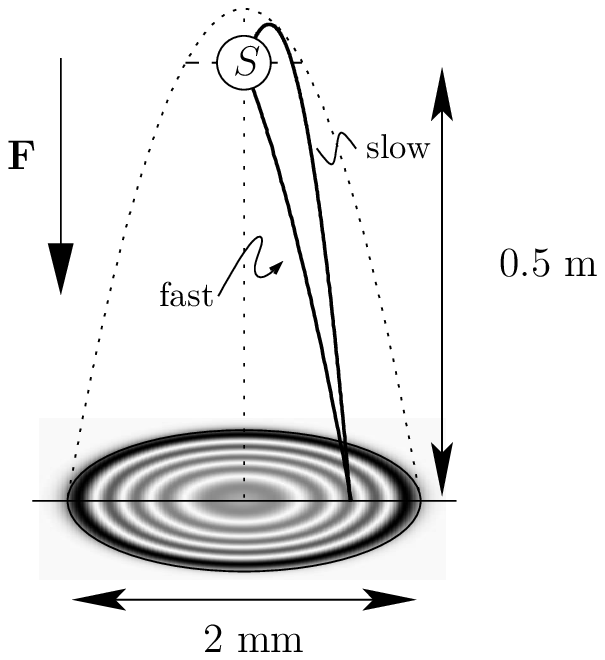}
\includegraphics[width=0.4\textwidth]{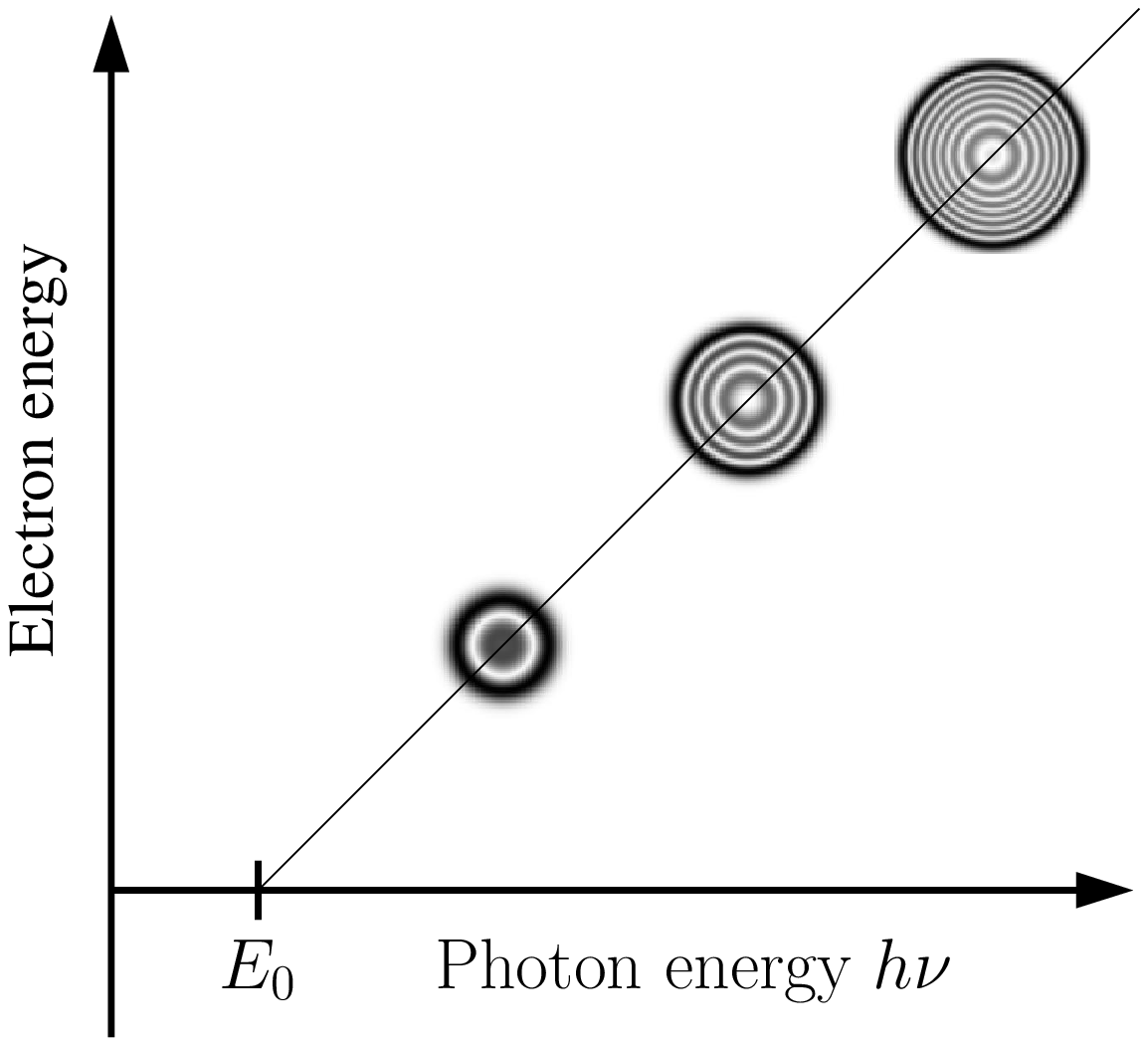}
\end{center}
\caption{Near--threshold detachment of oxygen ions: $\mathrm{O}^- \rightarrow \mathrm{O}+e^-$ in the presence of a homogeneous electric force ield $\mathbf{F}=e \mathbf{E}$. The two possible classical trajectories for a photoelectron leading from the source $\bigcirc\!\!\!\!\!S$ \, to any destination will give rise to interference on a distant detector screen. The fringe pattern in the current distribution depends sensitively on the energy. By counting the number of fringes the binding energy $E_0$ of the outer electron can be determined from Einstein's law \protect\cite{Blondel1996a,Blondel1999a,Blondel2005a,Bracher2003a}.}
\label{fig:blondel}
\end{figure}
Applying this photoelectric formula to negative ions in an applied electric field gives a new way to measure the kinetic energy of the detached electron, and therefore the binding energy of the negative ion, with unprecedented precision. 
(Even today it is difficult to calculate \textit{ab initio} the binding energy of the excess electron $E_0$ attached to a negative ion \cite{Andersen2004a}.)

The propagation of photoelectrons in a homogeneous external electric field has been experimentally studied by Blondel et al.\ \cite{Blondel1996a,Blondel1999a,Blondel2005a}. The left--hand side in Fig.~\ref{fig:blondel} illustrates the motion of a photoelectron subject to the electric force. The relevant Green function is that of a particle falling freely in a constant field \cite{Demkov1982a,Bracher1998a,Bracher2003a}. Two classical trajectories lead from the source (the negative ion) to any point on the detector. As in Young's double--slit experiment, an interference pattern is produced by the electron waves that travel along these two paths. From the interference pattern one can determine the kinetic energy of the electrons and plot it against the photon energy to check Einstein's law (\ref{eq:wf}) (right panel of Fig.~\ref{fig:blondel}), and to determine the binding energy $E_0$ of the electron. Experimental results are reported in Ref.~\cite{Blondel1999a}, Fig.~6. The recorded circular intensity fringes compare very well with the theoretical prediction \cite{Kramer2002a}. They show a highly accurate verification of Einstein's law which can be used to obtain the binding energy of O$^-$ with unprecedented accuracy \cite{Blondel1999a,Blondel2005a}.

Other new phenomena also appear in these experiments.  For example a static electric field opens up a sub--threshold ($E<0$) tunneling regime which also has been confirmed by experiment \cite{Gibson2001a}. An external magnetic field in combination with a crossed electric field can force the electron to stay in its initial bound state due to the lack of available final states \cite{Kramer2003a,Kramer2003d}. Experimentally, a suppression of the photocurrent has been observed \cite{Yukich2003a}.

The photoelectric effect in neutral atoms subject to external fields reveals similarly intriguing interference phenomena \cite{Nicole2002a}.  Their interpretation has led to a new understanding of classical periodic orbits, and their bifurcation and proliferation as order changes to chaos \cite{Du1988b,Gao1992a,Peters1993a,Main1994a,Haggerty1998a,Freund2002a}.  Finally, we mention a recent prediction for photoionization of a neutral atom in parallel electric and magnetic fields:  If the laser pulse is short, the released electrons will arrive at a detector in a chaotic pulse train \cite{Mitchell2004a}.

\section{Conclusions}

Einstein's theory of the photoelectric effect opened a door leading to the quantum theory of matter and radiation.  A century later, the photoelectric effect still provides new insights into the wave-particle duality, and it provides new ways to measure the structure and spectra of atoms and ions.

\section*{Acknowledgments}

This work was supported by the National Science Foundation and by
the DFG Project No. Kl 315/6.

\end{document}